\documentclass[useAMS,usenatbib,twocolumn,superscriptaddress,reprint,showpacs]{revtex4}

\usepackage{graphicx} 
\usepackage{amssymb} 
\usepackage{amsmath}

\usepackage{graphicx} 
\usepackage{amssymb}
\usepackage{amsmath}

\begin{document}

\title{Relic density and CMB constraints on dark matter annihilation with Sommerfeld
  enhancement}
\author{Jes\'us Zavala}
\email{jesus@mpa-garching.mpg.de}
\affiliation{Max-Planck-Institut f\"{u}r Astrophysik, Karl-Schwarzschild-Stra\ss{}e 1, 85740 Garching
bei M\"{u}nchen, Germany}
\author{Mark Vogelsberger}
\affiliation{Max-Planck-Institut f\"{u}r Astrophysik, Karl-Schwarzschild-Stra\ss{}e 1, 85740 Garching
bei M\"{u}nchen, Germany}
\affiliation{Harvard-Smithsonian Center for Astrophysics, 60 Garden Street, Cambridge, MA 02138, USA}
\author{Simon D. M. White}
\affiliation{Max-Planck-Institut f\"{u}r Astrophysik, Karl-Schwarzschild-Stra\ss{}e 1, 85740 Garching
bei M\"{u}nchen, Germany}

\begin{abstract}
We calculate how the relic density of dark matter particles is altered when
their annihilation is enhanced by the
Sommerfeld mechanism due to a Yukawa interaction between the annihilating
particles. Maintaining a dark matter abundance 
consistent with current observational bounds requires the normalization of the s-wave
annihilation cross section to be decreased compared to a model without
enhancement. The level of suppression depends on the specific parameters
of the particle model, with the kinetic decoupling temperature having the most
effect. We find that the cross section can be reduced by as much as
an order of magnitude for extreme cases. We also compute
the $\mu-$type distortion of the CMB energy spectrum caused by
energy injection from such Sommerfeld-enhanced annihilation. Our
results indicate that in the vicinity of resonances,
associated with bound states, distortions can be large enough to be excluded
by the upper limit $\vert\mu\vert\leq9.0\times10^{-5}$ found by the COBE/FIRAS experiment.
\end{abstract}


\maketitle








\section{Introduction}

If dark matter annihilates, the byproducts of the annihilation (positrons,
neutrinos, gamma-rays, etc.) can leave non-gravitational signatures that, if
observed, would be crucial for clarifying the nature of dark matter.

In recent years, a number of observations have highlighted anomalies that might
be explained by invoking dark matter annihilation in our Galactic halo. Among
them are: (i)
an anomalous abundance of positrons in cosmic rays above $10$~GeV according to
the PAMELA experiment \cite{Adriani-09} confirming and extending previous
measurements by experiments such as AMS-01 \cite{-07}; (ii) an excess of microwave emission from the
galactic center as measured by the WMAP experiment, known as the ``WMAP haze''
\cite{Hooper-Finkbeiner-Dobler-07} (iii) the apparent excess of diffuse
galactic gamma-rays with energies above 1GeV inferred from observations by
the EGRET satellite \cite{deBoer-05}; (iv) analyses on the balloon experiments ATIC and PPB-BETS
\cite{Chang-08,Torii-08} have reported an excess in the total flux of
electrons and positrons in cosmic rays. For the last two anomalies, we note that
recent observations by FERMI seem inconsistent with the claimed gamma-ray
excess in the EGRET data \cite{Porter--09}, and that the excess in
the electron and positron flux is smaller than previously thought
\cite{Abdo-09}. The latter is
actually consistent with a modified cosmic ray propagation model that does not
require additional primary sources of electrons and positrons \cite{Grasso-09}.

Although other astrophysical sources could explain these anomalies (see for
example \cite{Hooper-Blasi-DarioSerpico-09,Yuksel-Kistler-Stanev-09,Profumo-08} for an explanation for the
PAMELA excess based on particle acceleration by pulsars, and \cite{Fujita-09,Shaviv-Nakar-Piran-09}
for one based on supernova remnants), dark matter annihilation offers an attractive
solution. Large annihilation rates are needed, however, to explain the observations. In
particular for the PAMELA data, the required annihilation rate is typically a few orders of
magnitude larger than the value obtained by assuming the standard cross
section inferred from the observed abundance of dark matter together with a smooth
local distribution of dark matter \cite{Cirelli-09}. Thus, an additional
hypothesis is needed to boost the annihilation rate to the required levels
and this must not change the present-day abundance of dark matter. 

Such a boost is difficult to obtain from the effects of
substructures in the local dark matter distribution. Recent numerical simulations
predict this to be remarkably smooth \cite{Vogelsberger-09}. A
detailed analysis of the impact of substructure on the production of positrons
by \cite{Lavalle-08} came to a similar conclusion. 
The possibility of a nearby ``spike'' of dark matter produced by an
intermediate mass black hole seems also a priori implausible \cite{Bringmann-Lavalle-Salati-09}.

An alternative that has produced a plethora of papers in recent years is
that of a Sommerfeld enhancement to the cross section 
produced by the mutual interaction of the annihilating dark matter particles.
Such an interaction could be produced by a force carrier which might be any of the
standard model weak force gauge bosons \cite{Lattanzi-Silk-09} or a new force
carrier \cite{Arkani-Hamed-09}. For certain values of the parameters of these
models, the enhancement is easily large enough to boost the cross section to
the required values.

However, a large cross section has a significant impact on
other observables and may violate other constraints. For instance,
\cite{Kamionkowski-Profumo-08} showed that for the case where
the cross section increases as $1/v$ (a particular
case of Sommerfeld-enhancement models), where $v$ is the relative velocity of
the annihilating particles, there are severe constraints from measurements of
the diffuse extragalactic gamma-ray background radiation and from CMB constraints on
ionization and heating of the intergalactic medium (IGM) by annihilation
in the first generation of halos. In this scenario, the
boost factors required to fit the above anomalies would be
inconsistent with current constraints. This problem is avoided, however, by
more general cases of the Sommerfeld enhancement where the effect saturates at
low velocities \cite{Arkani-Hamed-09,Lattanzi-Silk-09}.

Recently, \cite{Galli-09} and \cite{Slatyer-Padmanabhan-Finkbeiner-09}
analyzed constraints on the annihilation cross section from
perturbations to the CMB angular power spectra resulting from heating and ionization
of the photon-baryon plasma at recombination. They found interesting upper
limits to models with Sommerfeld enhancement (see fig. 5 of \cite{Galli-09})
that already rule out some extreme cases.

At higher redshifts, the effects of the Sommerfeld
enhancement have scarcely been treated. For example,the annihilation of
dark matter impacts the predictions from big-bang
nucleosynthesis on the abundance of light elements
(e.g. \cite{Jedamzik-04}). This could put constraints on certain models with
Sommerfeld enhancement. However, this has only been studied in passing \cite{Hisano-09,Jedamzik-Pospelov-09}.

Also, the abundance of dark
matter today is commonly assumed to be unaltered by this effect, apparently because the
typical dark matter particle velocities at freeze-out are very large making the
enhancement very close to one at that epoch
\cite{Arkani-Hamed-09,Kuhlen-Madau-Silk-09}. As we will show, this reasoning
is flawed. 

The thermodynamic equilibrium between matter and radiation in the early
Universe would be perturbed by energy released during a certain process,
dark matter annihilation for example. This equilibrium tends to be restored
by different interaction mechanisms: Compton scattering, double Compton
emission and bremsstrahlung radiation. The efficiency of these to fully
restore equilibrium varies with redshift. For $z\gtrsim2\times10^6$ they are
efficient enough to restore distortions in the energy spectrum and thus, the
photon distribution is that of a black body with a slightly higher temperature
than the one in the case of no energy injection. For lower redshifts these
mechanisms can not restore the black body spectrum. In particular, for
$5.1\times10^4\lesssim z \lesssim2\times10^6$, the spectrum is perturbed into
a Bose-Einstein distribution with a chemical potential $\mu$
\cite{Illarionov-Siuniaev-75}.

In this paper, we revisit the
impact of the Sommerfeld enhancement on the relic particle abundance and
we show that its effect is not negligible. 
We also study, for the first time,
$\mu-$type distortions of the CMB spectrum due to energy deposition 
by dark matter annihilation in models with Sommerfeld enhancement. 

The paper is organized as follows. In section 2 we summarize the Sommerfeld
enhancement and describe how we include it in our calculations. The relic
density calculation is set out in detail in section 3. In section 4, the $\mu-$type
distortion to the CMB from annihilation with Sommerfeld
enhancement is calculated. Finally we present a summary and our conclusions in section 5.

\section{The Sommerfeld enhancement}

The Sommerfeld enhancement of dark matter annihilation is a nonrelativistic
quantum effect occurring when
annihilating particles interact through a potential of some sort. If their
kinetic energy is low enough, their
wave function is distorted, and a significant
enhancement (or suppression if the force is repulsive) to the annihilation
cross section occurs \cite{Hisano-Matsumoto-Nojiri-04,Arkani-Hamed-09,Lattanzi-Silk-09}. The product of
the relative velocity times the annihilation
cross section will therefore be boosted: $\sigma v=S(\sigma v)_0$, where
$(\sigma v)_0$ is the standard product of the tree level cross section times
the relative velocity and $S$ is the so called Sommerfeld boost. 

The simplest interaction is that for an attractive Yukawa potential
with coupling strength $\alpha_c$ mediated by a scalar boson of mass
$m_{\phi}$ (more complicated models exist where there are multiple force
carriers \cite{Arkani-Hamed-09}). In this case, the Sommerfeld
enhancement can be computed by solving the radial Schr\"odinger equation for
s-wave annihilation in the non-relativistic limit (we use natural units $\hbar=c=1$):
\begin{equation}\label{Sch}
\frac{1}{m_{\chi}}\frac{d^2\Psi(r)}{dr^2}+\frac{\alpha_c}{r}e^{-m_{\phi}r}\Psi(r)=-m_{\chi}\beta^2\Psi(r)
\end{equation}
where $m_{\chi}$ and $\beta$ are the mass and velocity of the dark matter
particle, respectively, and $\Psi(r)$ is the reduced two-body wave
function. Eq.~(\ref{Sch}) should be solved with the boundary condition:
$d\Psi/dr=im_{\chi}\beta\Psi$ as $r\rightarrow\infty$. The Sommerfeld boost
$S$ is simply given by:
\begin{equation}\label{S_0}
S=\frac{\left\vert\Psi(\infty)\right\vert^2}{\left\vert\Psi(0)\right\vert^2}
\end{equation}
In the limit of a massless force carrier ($m_{\phi}\rightarrow0$), the
potential reduces to a Coulomb interaction and Eq.~(\ref{Sch}) can be solved
analytically yielding a value of $S$ that is independent of the dark matter particle mass:
\begin{equation}\label{S_1}
S=\frac{\pi\alpha_{c}}{\beta}\left(1-e^{-\pi\alpha_{c}/\beta}\right)^{-1}
\end{equation}
In this limit, and for small velocities ($\beta\ll\pi\alpha_c$): $S\sim\pi\alpha_c/\beta$.
This is the reason why the Sommerfeld enhancement is usually associated with a ``$1/v$''
enhancement. 

In the more general case, where $m_{\phi}\neq0$, the ``$1/v$''
behavior is no longer valid for very small velocities because the
Yukawa interaction has a finite range. This can also be shown by expanding the
exponential term in Eq.~(\ref{Sch}) in powers of $m_{\phi}r$, then
the necessary condition to recover the Coulomb-like interaction
solution (Eq.~\ref{S_1}) is $\beta^2\gg\alpha_cm_{\phi}/m_{\chi}$
\cite{Lattanzi-Silk-09}, which, for fixed values of $\alpha_c$ and
$m_{\phi}/m_{\chi}$, will not be fulfilled for arbitrarily low
velocities. In the opposite regime, for $\beta^2\ll\alpha_cm_{\phi}/m_{\chi}$, a series of resonances
appear for specific values of $\alpha_cm_{\chi}/m_{\phi}$. These resonances
are associated with bound states, the so-called WIMPonium
\cite{March-Russell-West-09,Shepherd-Tait-Zaharijas-09}. In them, the enhancement
greatly increases with smaller relative velocity, as $S\sim1/\beta^2$. 
For even lower velocities and for fixed values of $\alpha_c$ and
$m_{\phi}/m_{\chi}$, the enhancement saturates
reaching a value that can be estimated by solving
Eq.~(\ref{Sch}) with $\beta=0$. Away from the resonances, an order of magnitude estimate of this value
is $S_{max}\sim6\alpha_cm_{\chi}/m_{\phi}$, reached for a threshold velocity of
$\beta\sim0.5m_{\phi}/m_{\chi}$ \cite{Lattanzi-Silk-09}. 

On the other hand, for large velocities ($\beta\gg\pi\alpha_c$),
the Coulomb approximation is valid and the enhancement is effectively reduced
to values very close to 1.

To include in detail the influence of the Sommerfeld enhancement in our study
we need to to solve numerically the Schr\"odinger equation. 

\section{Relic density calculation}

The evolution of the phase space distribution per particle of a given species
is given by the Boltzmann equation. For Majorana fermions that are stable, for
example, neutralinos if
they are the Lightest Supersymmetric Particle, the Boltzmann equation is \cite[e.g.][]{Gondolo-Gelmini-91}:
\begin{equation}\label{boltzmann}
\frac{{\rm d}n_{\chi}}{{\rm d}t}+3Hn_{\chi}=-\left<\sigma v\right>\left(n_{\chi}^2-\left(n_{\chi}^{EQ}\right)^2\right)
\end{equation}
where $H=\dot{a}/a$ is the Hubble parameter, $\left<\sigma v\right>$ is
the thermally averaged product of the total pair annihilation cross section times
the M\o ller velocity and {\small $n_{\chi}^{EQ}$} is the equilibrium dark
matter number density. In the remainder of the paper
we assume that dark matter is made of neutralinos. 

We solve Eq.~(\ref{boltzmann}) by using the standard freeze-out
approximation \cite[e.g.][]{Gondolo-Gelmini-91}. The freeze-out epoch is
defined as the time where the annihilation rate is equal to
the expansion rate of the Universe ($\Gamma\sim H$). In the standard case
where $\left<\sigma v\right>$ is velocity independent,
the annihilation process becomes subdominant after freeze-out(chemical decoupling) and the
``relic'' abundance of particles is roughly fixed\footnote{Annihilation events continue
  to occur after freeze-out, but they do not change the relic abundance significantly.}.

We define $\Delta=X-X_{EQ}$, where $X$ and $X_{EQ}$ are the comoving number
density out of and at equilibrium, then we can write Eq.~(\ref{boltzmann})
as:
\begin{equation}\label{boltzmann2}
\frac{{\rm d}\Delta}{{\rm d}t}+\frac{{\rm d}X_{EQ}}{{\rm d}t}=-a^{-3}\left<\sigma v\right>\Delta\left(2X_{EQ}+\Delta\right)
\end{equation}
Since before freeze-out the number density follows closely the equilibrium
solution then: ${\rm d}\Delta/{\rm d}t\ll {\rm d}X_{EQ}/{\rm d}t$, and at freeze-out:
$\Delta(t_f)\backsimeq cX_{EQ}(t_f)$, where $c$ is a constant of order
unity\footnote{The specific value of the parameter $c$ influences the results
  found on this paper. However, for a reasonable range of values:
  $c\in(0.1,3.0)$, the changes are in the percent level. We have adopted 
  $c=1.0$ in the results reported here.}. 
Under these approximations, the value of the freeze-out time can be
found by solving the following equation:
\begin{equation}\label{freeze_out}
\left.\frac{{\rm d}{\rm ln}X_{EQ}}{{\rm d}t}\right\vert_{t_f}\backsimeq
-a^{-3}\left<\sigma v\right>_f c(2+c)X_{EQ}(t_f)
\end{equation}

It is convenient to change the variable $X$ to the ratio $Y=n_{\chi}/s$ of the number
density to the total entropy density of the Universe. The entropy is dominated
by the contribution of relativistic particles {\small $s=(2\pi^2/45) g_{\ast
    S}T^3$}, where $g_{\ast S}$ are the effective degrees of freedom for the
entropy density, and $T$ is the radiation temperature.
In thermal equilibrium, the entropy per comoving volume is conserved
$sa^3=s_0\sim2918~{\rm cm}^{-3}$ (taking $T_0=2.728^{\circ}$K and $g_{\ast S}(T_0)=3.915$). 

It is also useful to change the time variable to
$x=m/T$ using the equation: {\small $t=0.301g_{\ast}^{-1/2}m_{Pl}(x/m)^2\equiv
  G_{\ast}(m,T)x^2/2$},
where $m_{Pl}=1.221\times10^{19}$GeV is the Planck mass and 
$g_{\ast}$ are the effective degrees of freedom for the total energy
density of the Universe, which are generally a function of temperature. We
have taken into account the temperature dependence of $g_{\ast}$ in our
calculations, whose values vary from $g_{\ast}\sim100$ at $T\sim100$~GeV to
$g_{\ast}\sim3$ at $T\sim10$~keV. This equation is valid during the radiation-dominated era. 
If we neglect the variations in $g_{\ast}(T)$ around the
freeze-out temperature\footnote{In the expected range of freeze-out
  temperatures for massive relics, $x\in(20,30)$, the function $g_{\ast}$ is roughly in a
  constant plateau, see for example fig. 1 of \cite{Gondolo-Gelmini-91}. This
  expected range is motivated by the values obtained in the standard
  calculations where $\left<\sigma v\right>=~$cte.}, we
can finally write Eq.~(\ref{freeze_out}) as:
\begin{equation}\label{freeze_out2}
\left.\frac{{\rm d}{\rm ln}Y_{EQ}}{{\rm d}t}\right\vert_{x_f}\backsimeq
-G_{\ast}(m,T_f)s_f\left<\sigma v\right>_f c(2+c)Y_{EQ}(x_f)x_f
\end{equation}
In the non-relativistic limit, the equilibrium solution takes a
simple form: {\small $Y_{EQ}(x)=0.145(g/g_{\ast S})x^{3/2}e^{-x}$}, where $g$
is the degeneracy factor for the particle species ($g=2$ for neutralinos). Therefore,
$x_f$ is simply given by the solution to the implicit equation:
\begin{equation}\label{freeze_out3}
e^{x_f}\backsimeq0.038\left(\frac{g}{g_{\ast}^{1/2}(T_f)}\right)m_{Pl}~c(2+c)\left(\frac{m_{\chi}}{x_f^2}\right)\left<\sigma v\right>_f
\end{equation}
where we have used the approximation $3x_f^{-1}/2\ll 1$, since we expect $x_{f}\gtrsim20$.

In the non-relativistic limit, the thermal average $\left<\sigma v\right>$ 
reduces to an average over a Maxwell-Boltzmann distribution function \cite{Gondolo-Gelmini-91}:
\begin{equation}\label{cross_nr}
\left<\sigma v\right>=\frac{x^{3/2}}{2\pi^{1/2}}\int_0^1 (\sigma
v)~\beta^2e^{-x\beta^2/4}d\beta
\end{equation}
where $v$ is the relative velocity of the annihilating particles. The expansion of $(\sigma v)$ in
powers of $\beta^2$ leads to the usual expansion of the non-relativistic
thermal average in powers of $x$, where the first term of the expansion
corresponds to the constant s-wave annihilation term $\left<\sigma
v\right>_S$. For the case when the cross section
is enhanced by the Sommerfeld mechanism the s-wave annihilation thermal
average is given by:
\begin{equation}\label{cross_nr_2}
\left<\sigma v\right>=\left<\sigma v\right>_S\left(\frac{x^{3/2}}{2\pi^{1/2}}\int_0^1 S(\beta)~\beta^2e^{-x\beta^2/4}d\beta\right)=\left<\sigma v\right>_S\mathcal{S}(x)
\end{equation}
where $\mathcal{S}(x)$ is the thermally averaged Sommerfeld enhancement.
The thermal average maintains the qualitative behavior of $S$ with velocity.
For example, when $S(\beta)\propto\beta^{-1}$, $\mathcal{S}(x)\propto
x^{1/2}\propto\sigma_v^{-1}$, where $\sigma_v$ is the velocity dispersion of
the dark matter particles \cite{Bovy-09}. We
have omitted in the notation of $\mathcal{S}(x)$ the dependence on the parameters of the
Yukawa interaction; in general $\mathcal{S}\equiv\mathcal{S}(x,\alpha_c,m_{\phi}/m_{\chi})$.

At freeze-out, although the dark matter particles are already
non-relativistic, the typical velocities are still very large, and we can
safely assume $S(\beta)\sim1$, i.e., $\left<\sigma v\right>_f \sim\left<\sigma
v\right>_S$. We have checked that this is indeed a good approximation
by solving fully the Schr\"odinger
equation, there is no significant change in the results that follow.  

The dark matter relic density can be calculated by solving the Boltzmann equation in the late times
regime ($t>t_f$) where the number density at equilibrium
$n_{\chi}^{EQ}$ is considerably lower than the out-of equilibrium solution and
can therefore be dropped from the Boltzmann equation. Thus, Eq.~(\ref{boltzmann}) can be
written in terms of the ratio $Y=n_\chi/s$ and the variable $x=m_{\chi}/T$ as 
\cite[e.g.][]{Gondolo-Gelmini-91}: 
\begin{equation}\label{boltzmann_late}
\frac{1}{Y(x_0)}=\frac{1}{Y(x_f)}+\sqrt{\frac{\pi}{45}}m_{\chi}m_{Pl}\left<\sigma
v\right>_S\int_{x_f}^{x_0}\frac{g_{\ast}^{1/2}(x)\mathcal{S}(x)}{x^2}dx
\end{equation}
where we have used Eq.~(\ref{cross_nr_2}), and the limits of integration corresponding to freeze-out $x_f=m_{\chi}/T_f$
and present-day values $x_0=m_{\chi}/T_{0}$.
The value of $Y(x_0)$ is connected to the value of the
ratio of the dark matter density to the critical density today
($\Omega_{\chi,0}=m_{\chi}n_{\chi,0}/\rho_{crit,0}$):
\begin{eqnarray}\label{Omega_Y0}
\Omega_{\chi,0}h^2=\frac{8\pi}{3M_{Pl}^2(H_0/h)^2}m_{\chi}s(x_0)Y(x_0)\nonumber\\
\sim2.757\times10^8\left(\frac{m_{\chi}}{GeV}\right)Y(x_0)
\end{eqnarray}

The integral in the r.h.s of Eq.~(\ref{boltzmann_late}) can be divided in
two. Before kinetic decoupling, the dark matter particles are still coupled
to the photon-baryon plasma through scattering with standard model particles
and therefore we can treat the temperatures of the radiation and of the neutralinos
interchangeably ($T_{\chi}=T$) for $x<x_{kd}=m_{\chi}/T_{kd}$, where $T_{kd}$ is the
temperature at kinetic decoupling.

After kinetic decoupling, the
temperature of the dark matter particles drops as $a^{-2}$ instead of
$a^{-1}$, as the temperature of radiation does. Thus, for this second part of
the integral in Eq.~(\ref{boltzmann_late}), one must be
careful in the definition of the variable $x$ in the thermal average (Eq.~\ref{cross_nr_2})
and in the Boltzmann equation (Eq.~\ref{boltzmann_late}). The thermal average uses a
Maxwell-Boltzmann distribution function with the temperature of the dark
matter ``gas'', whereas Eq.~(\ref{boltzmann_late}) was written as a
function of $x=m_{\chi}/T$, where $T$ is the radiation temperature. We can
easily account for this by evaluating $S(x_{\chi})$ in Eq.~(\ref{cross_nr_2})
as a function of the radiation temperature by using the relation:
$x_{\chi}=x^2/x_{kd}$. As pointed out by \cite{Dent-Dutta-Scherrer-09}, in
this regime ($x>x_{kd}$), for the case where $S(x_{\chi})\propto
x_{\chi}^{1/2}\propto x$, the relic density decays logarithmically:
$Y\propto 1/{\rm ln}x$. In section 2 we mentioned that for particular values
of $\alpha_c$ and $m_{\phi}/m_{\chi}$, the Sommerfeld enhancement has
resonances that for very low velocities produce very large values of $S$, in
these cases, $S(x_{\chi})\propto x_{\chi}\propto x^2$, thus, the relic density
decreases as $1/x$.

However, these large suppressions of the relic density are eventually cut-off
by the saturation of $S$, which occurs before matter-radiation
equality. Consider, for example, a $100$ GeV neutralino. Since
$z_E\sim4\times10^3$, and because $T_{\chi}\sim(z_E/z_{kd})^2$ with
$z_{kd}\sim4\times10^{10}$ (the exact value of $z_{kd}$ depends on the
specific supersymmetric model, this is a typical value for a neutralino of
this mass), we have a dark matter velocity dispersion of the order of
$10^{-8}$ by $z=z_E$, which is sufficiently low to reach saturation,
unless the combination of $\alpha_c$ and $m_{\phi}/m_{\chi}$ is fine-tuned to
be close to a resonance.

In summary, a proper calculation of the relic density needs to include the
effects of kinetic decoupling. We do so by choosing a model with neutralino
dark matter, and by using the results of \cite{Bringmann-09}, who
made a scan of the parameter space in mSUGRA (minimal supergravity) and more
general MSSM models, and give a range for the values of the kinetic decoupling temperature as a
function of neutralino mass (see fig. 2 of their paper). Typical values 
for $m_{\chi}\in(100,5\times10^3)$~GeV are in
the range $x_{kd}\in(200,2\times10^4)$.

\begin{figure}
\centering
\includegraphics[height=8.5cm,width=10.0cm]{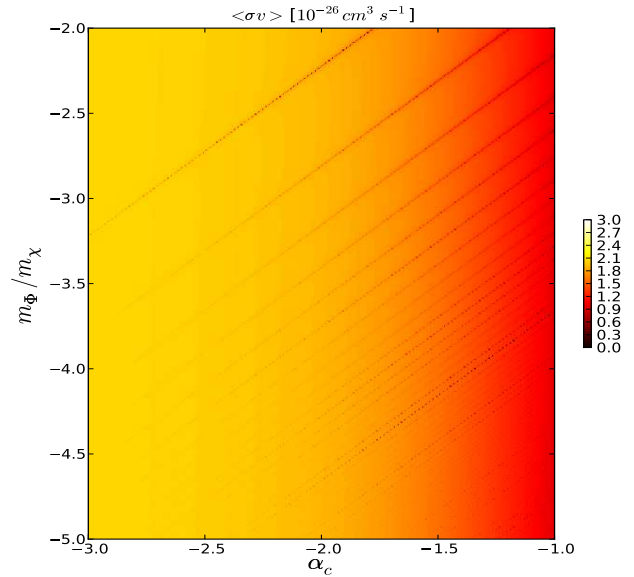}
\caption{Values of the normalization of the thermally averaged cross section $\left<\sigma
  v\right>_S$ that give the proper relic density $\Omega_{\chi}h^2=0.1143$ in
  the parameter space ($m_{\phi}/m_{\chi}$, $\alpha_c$) for
  $m_{\chi}=100~{\rm GeV}$ and $T_{kd}=8$~MeV. The values are given in
  a color scale shown on the right hand side in units of $1\times10^{-26}{\rm cm^3 s^{-1}}$.}
\label{cross_section}
\end{figure}

For a fixed value of $\Omega_{\chi}h^2$, we can solve Eqs.~(\ref{freeze_out3})
and (\ref{boltzmann_late}) simultaneously to get the values of $x_f$ and
$\left<\sigma v\right>_S$ that give the correct relic abundance. The
value of the present-day dark matter density according to the WMAP-5 year
data, at the $2\sigma$ level, is $\Omega_{\chi}h^2=0.1143\pm0.0068$
\cite{Komatsu-09}. Fig.~\ref{cross_section} shows the value of
$\left<\sigma v\right>_S$ consistent with $\Omega_{\chi}h^2=0.1143$
for a scan in the parameter space ($m_{\phi}/m_{\chi}$, $\alpha_c$) 
with $m_{\chi}=100~{\rm GeV}$ and $T_{kd}=8$~MeV. The values of the thermally averaged
cross section are color-coded in units of $1\times10^{-26}{\rm cm^3 s^{-1}}$.
For comparison, in the absence of any Sommerfeld enhancement the required
cross section is $\sim2.4\times10^{-26}{\rm cm^3 s^{-1}}$.

Since the typical velocities of the dark matter particles between $x_{f}$ and
$x_{kd}$ are still very high, the Coulomb regime is valid in this
range for most of the parameter space (recall that the Yukawa part of the
potential can be ignored when $\beta^2\gg\alpha_cm_{\phi}/m_{\chi}$), thus,
if we cut the integral in the r.h.s of Eq.~(\ref{boltzmann_late}) at $x=x_{kd}$,
the relic density calculation would depend only in the strength of the coupling
$\alpha_c$. This dependence can be seen in the overall trend from left to
right in Fig.~\ref{cross_section}. However, after kinetic decoupling, the WIMPs get colder much
faster than the radiation and the Sommerfeld enhancement depends not only on
$\alpha_c$ but also on $m_{\phi}/m_{\chi}$. This is particularly significant
near resonances where the enhancement is large enough to suppress the dark
matter abundance by a factor of a few, therefore, the cross section
normalization needs to be suppressed by this same factor in order to be
consistent with the observed relic density. 

\begin{figure}
\centering
\includegraphics[height=8.5cm,width=10.0cm]{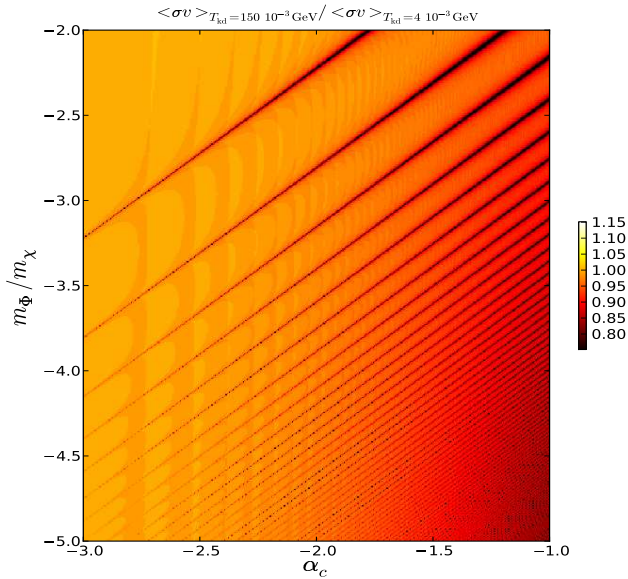}
\caption{Map ratio of the value of the cross section giving the correct relic
  density $\Omega_{\chi}h^2=0.1143$ with different kinetic decoupling
  temperatures, $T_{kd}=150$~MeV and $T_{kd}=4$~MeV for a $100$~GeV neutralino.}
\label{cross_section_ratio_Tkd}
\end{figure}

As shown by the discussion in the previous paragraph, the particular value of the
kinetic decoupling temperature has an impact on Fig.~\ref{cross_section}. The
closer this temperature is to the freeze-out temperature, the stronger the impact will be. To
appreciate this better, we show in Fig.~\ref{cross_section_ratio_Tkd}
the ratio of two maps analogous to the map in Fig.~\ref{cross_section} but for
two extreme values of the kinetic decoupling temperature for $m_{\chi}=100$~GeV: 
$T_{kd}=150$~MeV and $T_{kd}=4$~MeV. For large
values of the coupling strength and low values of $m_{\phi}/m_{\chi}$, and
particularly near resonances, the cross section at freeze-out needs to be lower for a higher
kinetic decoupling temperature to produce the correct relic density. From the figure we
see that uncertainties in the kinetic decoupling temperature introduce changes
to $\langle\sigma v\rangle_S$ at the most of order $30\%$ for a $100$ GeV neutralino.

Since current uncertainties on the abundance of dark matter are in the
percent level, the impact of the particular value of $\Omega_{\chi}h^2$ on
Fig.~\ref{cross_section} is not so large. By exploring changes in
$\Omega_{\chi}h^2$ within the $2\sigma$ bounds we found changes in the values
of $\langle\sigma v\rangle_S$ of $20\%$ at the most.

These results are also not very sensitive to changes in the neutralino
mass, because the relic density is nearly independent of it. This can be seen by looking at
Eqs.(\ref{boltzmann_late}-\ref{Omega_Y0}). If we neglect the term  $1/Y(x_f)$,
then, for a fixed value of $\langle\sigma v\rangle_S$, $\Omega_{\chi}h^2$
depends on the neutralino mass just through the integral in the r.h.s of
Eq.~(\ref{boltzmann_late}). Moreover, since the
Sommerfeld enhancement only depends on the neutralino mass through the
ratio $m_{\phi}/m_{\chi}$, a change on $m_{\chi}$ can always be compensated
by choosing a different mass for the force carrier. The remaining dependence
is through the values of $x_f$ and $x_{kd}$. 

Of course, to get the precise impact on the value  
of the cross section required for a particular particle model, it is
necessary to do the calculation for that model. Uncertainties on
the kinetic decoupling temperature and on the abundance of dark matter will
have a minor-to-mild impact on the quantitative results. However,
the general behavior and order of magnitude of the expected values of 
$\langle\sigma v\rangle_S$ can be directly read off from Fig.~\ref{cross_section}.

In summary, the relevant difference between the relic density calculation
for the standard case where $\langle\sigma v\rangle$ is constant and the
case we study here is given by the onset of the Sommerfeld enhancement after
kinetic decoupling. Before this phase, the velocities of dark matter particles
are too large for the enhancement to be relevant, but once the decoupling
happens the typical velocities decrease quickly since the temperature
of the dark matter fluid drops as $a^{-2}$. This causes a rapid increase on
the enhancement values and the annihilation process takes a relevant role once
more decreasing the dark matter density. On the contrary, the comoving density
remains constant after freeze-out for the standard case. The aforementioned
effect continues to be important until the enhancement saturates fixing the
value of $\langle\sigma v\rangle$ to a maximum. Afterwards, the annihilation
rate becomes subdominant compared to the expansion rate of the Universe and
thus, the comoving dark matter density remains fixed.

Owing to this difference, the value of the annihilation cross section before
the freeze-out epoch needs to be smaller for the case with Sommerfeld
enhancement than for the standard one in order to produce the observed dark
matter density. With a lower annihilation rate, the dark matter particles go
through chemical decoupling at earlier times, and therefore the dark matter
density is higher at these epoch than in the standard case. This higher value
compensates the subsequent decrease of the density due to the Sommerfeld
enhancement. In this way, the final relic density can be consistent with the
expected value for the abundance of dark matter today.

\section{$\mu$-type distortion of the CMB spectrum}

Measurements of the CMB spectrum by the FIRAS instrument onboard COBE have
shown that is nearly that of a perfect blackbody with temperature
$T_0\sim2.728^\circ$~K. Deviations from a pure black-body spectrum are, however,
expected if there is any energy input in the early Universe. If this
energy is injected between $5.1\times10^4\lesssim z\lesssim2\times10^6$
(12~eV$\lesssim T\lesssim$470~eV), the
spectral distortion is a Bose-Einstein $\mu-$type distortion where $\mu$ is
the chemical potential \cite{Illarionov-Siuniaev-75}. If the  number of
photons injected during the energy release is small, compared to the number of
photons in the radiation plasma, then
$\mu\sim1.4\delta\rho_{\gamma}/\rho_{\gamma}$, where $\delta\rho_{\gamma}$ is
the injected energy and $\rho_{\gamma}$ is the energy density of the CMB
photons. 

The current observational limit for the
$\mu$-type distortion at the $95\%$ confidence level is
$\vert\mu\vert\leq9.0\times10^{-5}$ \cite{Fixsen-96}. In principle, this limit
could already have been improved by nearly two orders of magnitude by now
given recent technological advances \cite{Fixsen-Mather-02}. 

In the case of energy deposited by annihilation of neutralinos, the $\mu-$type
distortion was studied by \cite{McDonald-Scherrer-Walker-01} for the case of s-wave and p-wave cross
sections. More recently it was also mentioned by \cite{Chluba-10}. The value of $\mu$ is given by:
\begin{equation}\label{mu_eq}
\mu=1.4\frac{\delta\rho_{\gamma}}{\rho_{\gamma}}=1.4\int_{t_1}^{t_2}\frac{\dot{\rho_{\gamma}}}{\rho_{\gamma}}dt=1.4\int_{t_1}^{t_2}\frac{fm_{\chi}\left<\sigma
  v\right>n_{\chi}^2}{\rho_{\gamma,0}a^{-4}}{\rm d}t
\end{equation}
where, $\rho_{\gamma,0}$ is the present-day energy density of the CMB
($\Omega_{\gamma}h^2\sim2.47\times10^{-5}$), $t_1-t_2$ is the time interval
corresponding to the energy injection and $f$ is the the efficiency in which
the injected energy is transformed into heat\footnote{In Eq.~(\ref{mu_eq}) we
  used that the rate of annihilation events per unit volume for neutralinos,
  being Majorana particles, is $fm_{\chi}/2$ and that the energy deposited in
  each event is $2fm_{\chi}$.}. In principle, $f$ depends on the channel of
annihilation and on time. However at the relevant redshifts ($z\gg1000$),
$f$ is basically given by the annihilation channel
\cite{Slatyer-Padmanabhan-Finkbeiner-09}. For electrons and photons $f\sim1$,
but for annihilation into $\tau$'s for example, some of the energy is lost in
neutrinos. Nevertheless, for all relevant channels,
\cite{Slatyer-Padmanabhan-Finkbeiner-09} found that $f>0.25$ for $z>2500$. We will adopt
$f=1$, our results can be reinterpreted easily for a different value of $f$. 

Since the relevant redshift range of energy injection is in the radiation
dominated era and after kinetic decoupling, we can write $t=t_{kd}(1+z_{kd})^2
/ (1+z)^2=t_{kd}x_{\chi}/x_{\chi}^{kd}$. It is also convenient to write Eq.~(\ref{mu_eq}) in terms of the 
variable $x_{\chi}=m_{\chi}/T_{\chi}$:
\begin{eqnarray}\label{mu_eq1}
  \mu=1.4f\left(\frac{\left<\sigma
    v\right>_S}{m_{\chi}\rho_{\gamma,0}}\right)\left(\frac{t_{kd}}{x_{kd}(1+z_{kd})^4}\right)\dot\nonumber\\
  \dot{}\int_{x_{\chi}^1}^{x_{\chi}^2}\mathcal{S}(x_{\chi})x_{\chi}^{-1}\rho_\chi^2(x_{\chi}){\rm d}x_{\chi} 
\end{eqnarray}
where:
\begin{equation}\label{x_i}
x_{\chi}^{1,2}=\frac{m_{\chi}T_{kd}}{T_0^2({1+z_{1,2}})^2}
\end{equation}

The density of neutralinos changes with time (temperature through the
variable $x_{\chi}$) according to the Boltzmann equation. Its value at a
given radiation temperature is related to the ratio $Y$ given by
Eq.~\ref{boltzmann_late} (recall that $\rho_\chi=m_\chi s Y$) by replacing
$x_0$ with the corresponding $x=m_\chi/T$ and using the relation
between the radiation temperature and that of the neutralino gas:
\begin{equation}
  \rho_\chi(x_{\chi})=m_\chi s(x=(x_{kd}x_\chi)^{1/2})Y(x=(x_{kd}x_\chi)^{1/2})
\end{equation}
Note that for the redshift range of interest $T<$470~eV and thus $g_{\ast S}$,
which enters in the value of the entropy $s(x)$, is a constant equal to
$g_{\ast S}(x_0)=3.915$.

\begin{figure}
\centering
\includegraphics[height=8.5cm,width=10.0cm]{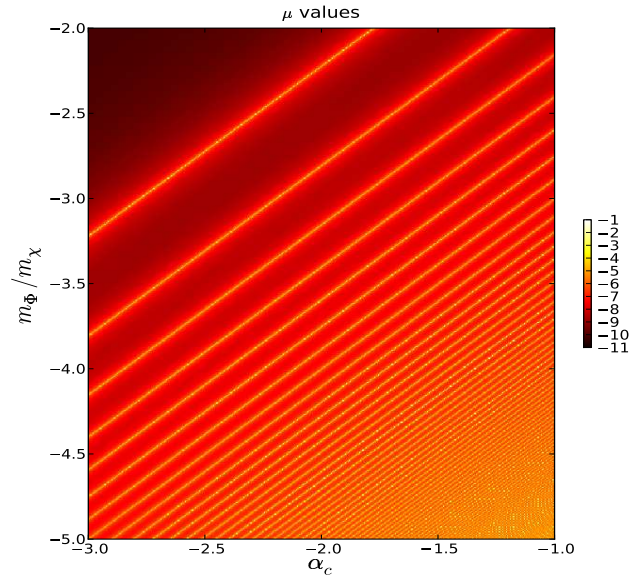}
\caption{Scan of the parameter space ($m_{\phi}/m_{\chi}$, $\alpha_c$) with
  the expected values of the $\mu$-type distortion to the CMB spectrum for
  $m_{\chi}=100$~GeV, $T_{kd}=8$~MeV and cross section values satisfying the
  constraint in the relic density: $\Omega_{\chi}h^2=0.1143$. The $2\sigma$
  observational upper limit on $\vert\mu\vert$ is $9\times10^{-5}$. The plot shows the
  values of the logarithm of $\mu$ color-coded according to the scale on the right.}
\label{mu}
\end{figure}

Now, $t_{kd}(1+z_{kd})^2=t_E(1+z_E)^2=t_0(1+z_E)^{1/2}$, where $z_E$ is the
redshift of matter and radiation equality and $t_0$ is the age of the Universe. In the last equality we have
used the relation between the scale factor and time for an Einstein-de Sitter
Universe. Since $(1+z_E)=\Omega_{\chi,0}/\Omega_{\gamma,0}$, we can finally
write:
\begin{eqnarray}\label{mu_eq2}
  \mu=1.4f\left(\frac{\Omega_{\chi,0}h^2}{\Omega_{\gamma,0}h^2}\right)^{3/2}\left(\frac{\Omega_{\chi,0}\rho_{crit,0}}{m_{\chi}}\right)t_{0}\left<\sigma
  v\right>_S\dot\nonumber\\
  \dot{} \int_{x_{\chi}^1}^{x_{\chi}^2}\mathcal{S}(x_{\chi})x_{\chi}^{-1}\left(\frac{Y(x=(x_{kd}x_\chi)^{1/2})}{Y(x_0)}\right)^2{\rm
    d}x_{\chi}
\end{eqnarray}
Note that the function in parentheses inside the integral is equal to one
if the comoving density of neutralinos is no longer evolving (has frozen) by
$z=2\times10^6$. This is the case in the majority of the parameter space of
the Yukawa interaction because by this redshift the velocity dispersion of
dark matter particles is low enough for the Sommerfeld enhancement to be
saturated. As we mentioned before, this means that the comoving density is
essentially frozen afterwards.

\begin{figure}
\centering
\includegraphics[height=8.5cm,width=8.5cm]{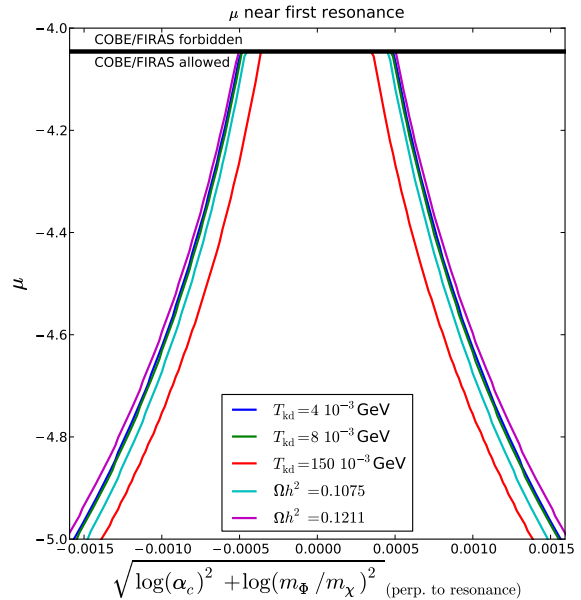}
\caption{The values of the $\mu$ distortion parameter near the first resonance
  in the upper left corner of Fig.~\ref{mu} as a function of the perpendicular
distance to the resonance. The lines with different colors correspond to
different values of $T_{kd}$ and $\Omega_{\chi}h^2$ as marked in the
legend. The solid black horizontal line shows the upper limit on $\mu$
according to the COBE/FIRAS experiment. As in Fig.~\ref{mu}, $m_{\chi}=100$~GeV.}
\label{mu_resonance}
\end{figure}

Fig.~\ref{mu} shows the values of $\mu$ found by solving Eq.~(\ref{mu_eq2}) for
the same scan of the parameter space ($m_{\phi}/m_{\chi}$, $\alpha_c$) as in
previous figures. We took as fiducial values 
$m_{\chi}=100~{\rm GeV}$, $T_{kd}=8$~MeV and $\Omega_{\chi}h^2=0.1143$, that is,
the values of the normalization to the cross section are required to give the observed dark matter
abundance. The values of
the logarithm of $\mu$ appear color-coded according to the scale in the right
of the figure. The resonances inherited from $\mathcal{S}(x)$ and shown in 
Fig.~\ref{cross_section} are also clear in this
figure. Most of the region on the lower right corner is above, or very close
to, the observational upper limit $\vert\mu\vert\leq9\times10^{-5}$. Points of
the parameter space very near resonances also give values of $\mu$ that can be
excluded. To show in more detail the behavior near a resonance, we present in 
Fig.~\ref{mu_resonance} a plot of the values of $\mu$ as a function of the
perpendicular distance, in parameter space, to the first resonance in the upper left of
Fig.~\ref{mu}. Lines with different colors show the magnitude of the
change produced by different values of the kinetic decoupling temperature and
of $\Omega_{\chi}h^2$, corresponding to the
legends in the figure. The upper limit on $\mu$ according to the
COBE/FIRAS experiment is also marked as a horizontal line. The
figure shows how close one can get to a resonance without violating the
current constraint. 

This is a significant point, because values very
close to a resonance are often invoked to provide a large boost and so to match observations
of the positron excess in cosmic rays.

According to Eq.~(\ref{mu_eq2}), the value of $\mu$ decreases with increasing
mass ($\mu\propto 1/m_{\chi}$) because $\langle\sigma v\rangle_S$ is nearly
independent of mass, and because, except very near
resonances, $S(x_{\chi})$ is already saturated by $z=2\times10^6$. In this
way, our results can be easily scaled to the desired neutralino mass. Care is
needed very close to the resonances since the simple scaling $\mu\propto
1/m_{\chi}$ is no longer valid there.

\begin{figure}
\centering
\includegraphics[height=8.5cm,width=8.5cm]{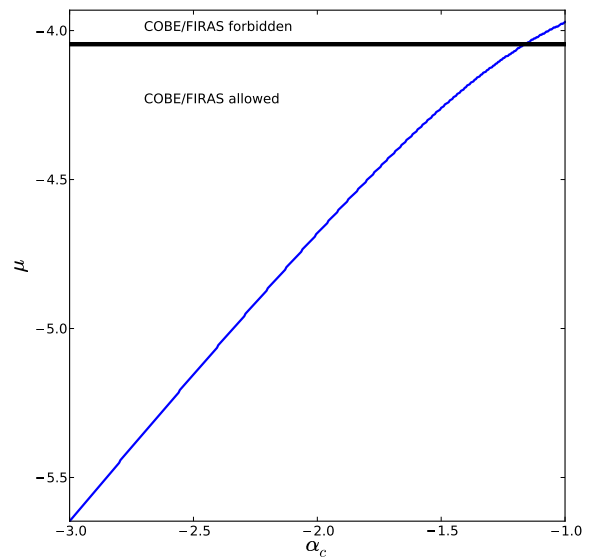}
\caption{The same as Fig.~\ref{mu} but for the case where the interaction is
strictly Coulomb-like. We have taken fiducial values of $m_{\chi}=100$~GeV,
$T_{kd}=8$~MeV and $\Omega_{\chi}h^2=0.1143$. Values above the black solid
horizontal line are excluded by the current limits of $\mu$ according to the
COBE/FIRAS experiment.}
\label{mu_coulomb}
\end{figure}

Finally, we show in Fig.~\ref{mu_coulomb} the analogous result for the values
of $\mu$ in the case where the interaction responsible of the Sommerfeld
enhancement is Coulomb-like ($m_{\phi}\rightarrow0$). In this case the enhancement
depends only on $\alpha_{c}$.  The plot shows, for the same fiducial case
($m_{\chi}=100$~GeV, $T_{kd}=8$~MeV and $\Omega_{\chi}h^2=0.1143$), that all cases with
$\alpha_c\gtrsim6\times10^{-2}$ are already excluded.

\section{Summary and Conclusions}

The prospects for dark matter detection have increased considerably in recent
years. There is a continual advance in the sensitivity of detectors on Earth
that look for direct dark matter elastic scattering with nuclei, and in
experiments that search indirectly for dark mater by looking for
non-gravitational signatures of the byproducts of its hypothetical 
annihilation. Such technological improvements may lead, in the near
future, to a definite proof of the existence of dark matter.

Perhaps dark matter has produced non-gravitational
signals that have already been detected. The recently reported excesses of positrons in cosmic rays by PAMELA 
and of electrons+positrons by FERMI/ATIC/PPB-BETS can be explained by dark
matter annihilation. Although other explanations with a different astrophysical
origin are also possible, a solution based on dark matter is an attractive
possibility. 

However, such solution seems to require an attractive force between the dark
matter particles to enhance their annihilation through the Sommerfeld
mechanism. For a Yukawa interaction via a single scalar, the magnitude of 
the enhancement depends on the coupling strength  of the interaction
$\alpha_c$, on the mass
ratio of the force carrier to the dark matter particle $m_{\phi}/m_{\chi}$ and
on the relative velocities of the annihilation particles $\beta$.

Several recent papers have computed the boost to the
annihilation due to a Sommerfeld enhancement
(in the Galactic halo and/or in its subhalos)
\cite{Lattanzi-Silk-09,Arkani-Hamed-09,Bovy-09}. One of the main
aims of these works is to show that for certain values of $\alpha_c$ and 
$m_{\phi}/m_{\chi}$, this mechanism is able to produce large enough boosts to
explain the cosmic ray anomalies.

Only a handful of studies have addressed in detail the impact that such an
enhancement has in the early Universe. In \cite{Kamionkowski-Profumo-08}, the authors found
that if the cross section increases with decreasing relative velocity as
$1/\beta$ (which is valid in a certain regime for general Sommerfeld
enhancement models), dark matter annihilation in the first halos
would heat and ionize the IGM, violating current constraints from the
CMB. As was later pointed out by \cite{Lattanzi-Silk-09} and
\cite{Arkani-Hamed-09}, this problem is alleviated in more general models
where the enhancement saturates at low velocities. In \cite{Galli-09} and 
\cite{Slatyer-Padmanabhan-Finkbeiner-09}, a constraint 
on the annihilation cross section was obtained by considering limits on the energy deposition by
annihilation at recombination. The constraint reported by
\cite{Slatyer-Padmanabhan-Finkbeiner-09} is 
$\langle\sigma v\rangle_{REC}<3.6\times10^{-24}{\rm cm}^3{\rm
  s}^{-1}(m_{\chi}/1{\rm TeV})/f_{REC}$, where $f_{REC}$ is an average
efficiency of energy injection into the IGM by annihilation at recombination. 

In the present paper, we have analyzed the impact of dark matter annihilation
with Sommerfeld enhancement at higher redshift.

In the first place, we have revisited the calculation of the dark matter
particle abundance by solving the Boltzmann equation from freeze-out through
the epoch of kinetic decoupling, including a full solution to the
Schr\"odinger equation.
Contrary to previous
claims \cite{Arkani-Hamed-09,Kuhlen-Madau-Silk-09}, we have found a significant
suppression of the relic density, in agreement with a recent work by
\cite{Dent-Dutta-Scherrer-09}. This suppression is particularly important
near resonances, which are typically invoked to explain the cosmic
ray anomalies. 

We found that to fit the observed dark matter abundance, 
the normalization of the cross section needs to be lowered by
up to a factor of 10 compared to the case without enhancement (see
Fig.~\ref{cross_section}). The result depends on the coupling strength and the
proximity to a resonance. Exploring a broad range of dark matter particle
masses and kinetic decoupling temperatures, we found a minor-to-medium impact of
these parameters on our results; variations on $T_{kd}$ have the strongest impact.

Secondly, we have calculated the amount of energy deposited by dark matter
annihilation into the radiation plasma in the redshift range
$5.1\times10^4<z<2\times 10^6$. Energy injection at this epoch would create a
distortion in the CMB energy spectrum first pointed out by
\cite{Illarionov-Siuniaev-75}. The COBE/FIRAS
experiment has put constraints on this Bose-Einstein $\mu-$type
distortion. The upper limit on the chemical potential associated to the
distortion is $\vert\mu\vert\leq9.0\times10^{-5}$ at the $2\sigma$ level
\cite{Fixsen-96}. We have found that very near resonances, annihilation with
Sommerfeld enhancement is already ruled out by this constraint on
$\mu$ (see Fig~\ref{mu}). Quantitatively, the ``safe'' range of proximity to a
resonance depends on the particular model, higher values of dark matter mass
and kinetic decoupling temperature allow a closer proximity to the resonance 
(see Fig.~\ref{mu_resonance}). In the case where the force carrier is
massless, the Yukawa interaction reduces to a Coulomb one. Our findings are
much more stringent in this case with values of $\alpha_c>6\times10^{-2}$
already ruled out (see Fig.~\ref{mu_coulomb}).

Improved upper limits on the $\mu$ parameter by a null distortion detection
in the CMB spectrum would rule out a larger
region of the parameter space for the Yukawa
interaction. In \cite{Fixsen-Mather-02}, it was pointed out that an improvement close to
two orders of magnitude has already been possible for a number of years, and 
\cite{Mather-07} suggests that another order of magnitude is perhaps within
reach. An upper limit on $\vert\mu\vert$ of the order of $10^{-7}$ would
certainly exclude large regions of the parameter space.
It would exclude to a large extent near-resonance regions, which are the ones
that produce the largest boosts to the annihilation.
On the other hand, the detection of a distortion could possibly
tell us something about the parameters of the Yukawa interaction and the nature
of the force carrier $\phi$.

In summary, our results indicate that for a given
set of parameters $(m_{\chi},T_{kd},m_{\phi},\alpha_c)$, it is necessary to
compute in detail the relic density to obtain the range of values for the
normalization to the cross section that are compatible with current estimates of
the dark matter abundance. Once this normalization is known, the energy
input producing a $\mu-$type distortion in the CMB spectrum can be
computed to check whether the particular model violates current constraints. 
Only for allowed models can a boost factor for local dark matter
annihilation be computed and advocated. 

Our findings show that the local
boosts reported in the literature need to be renormalized to the proper value
of $\langle\sigma v\rangle_S$ implied by the observed relic density. This
renormalization can exceed a factor of 10 in extreme cases. 

\begin{figure}
\centering
\includegraphics[height=8.5cm,width=10cm]{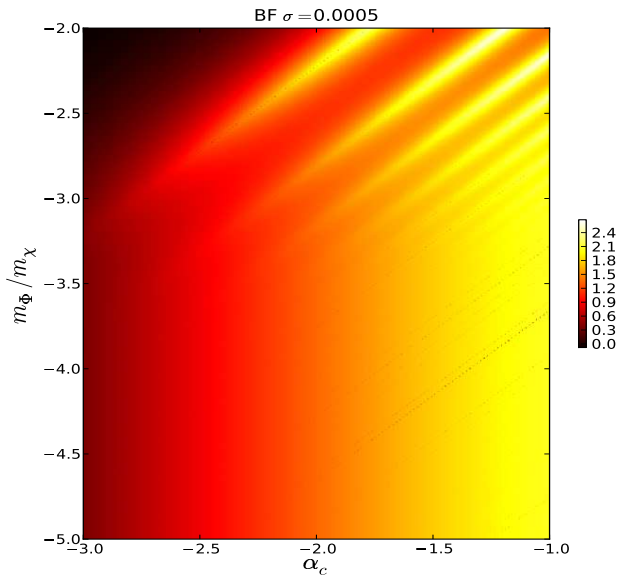}
\caption{The values of the boost factor, relative to $\langle\sigma
  v\rangle_S=3\times10^{-26}{\rm cm}^3{\rm s}^{-1}$ for the parameter scan of
  the Yukawa interaction. These values are obtained according to the cross
  section values of Fig.~\ref{cross_section}, which are consistent with
  $\Omega_{\chi}h^2=0.1143$. A Maxwell-Boltzmann velocity distribution
  with $\sigma_v=5\times10^{-4}$ for a neutralino with $m_{\chi}=100$GeV and
  $T_{kd}=8\times10^{-3}$GeV was used. The values are color-coded
  logarithmically according to the scale on the right.}
\label{boost_factor}
\end{figure}

A recent analysis of the PAMELA and FERMI data by
  \cite{Bergstrom-Edsjo-Zaharijas-09} suggests that boost factors,
  over a standard value of $\langle\sigma v\rangle_S=3\times10^{-26}{\rm
    cm}^3{\rm s}^{-1}$,
of order 1000 or higher are required to explain both datasets
simultaneously. We note that smaller boost factors ($\sim 100$) can still fit the
PAMELA data, but not the FERMI one, for a 100~GeV neutralino, which is the
standard case we have considered here. These large boosts are
also found in the analysis by \cite{Meade-09}. It has been argued in
the past that these boosts can be achieved only by invoking
Sommerfeld-enhanced annihilation. For instance, \cite{Arkani-Hamed-09} and
\cite{Bovy-09} obtain maximum boost factors $\sim1000$ by
assuming a local Maxwell-Boltzmann velocity distribution with a velocity dispersion of
$150~{\rm kms}^{-1}$ ($5\times10^{-4}c$), which roughly corresponds to the
estimated local dark matter velocity dispersion. We find that these boosts are modified
once the proper cross section is used. 

In Fig.~\ref{boost_factor} we show the boost factors, i.e., the multiplicative
factor to $\langle\sigma v\rangle_S=3\times10^{-26}{\rm cm}^3{\rm s}^{-1}$,
that we find for a Maxwell-Boltzmann velocity distribution with $\sigma_v=5\times10^{-4}$
for the parameter space of the Yukawa interaction. The figure is for
$m_{\chi}=100$GeV, $T_{kd}=8\times10^{-3}$GeV and is consistent with a relic
abundance $\Omega_{\chi}h^2=0.1143$. Keep in mind, however, that
changes in the neutralino mass have almost no impact
on the normalization of the cross section. Different kinetic decoupling
temperatures and variations on $\Omega_{\chi}h^2$ have also only a small effect
($<60\%$ in combination). Thus, Fig.~\ref{boost_factor} is approximately
correct for all the cases considered in this paper. The figure shows that
even in extreme cases, for resonances in the upper right of the figure,
the boost factors are $<500$. In more favored regions of the parameter
space $\alpha_c\gtrsim10^{-2}$, $m_{\phi}/m_{\chi}\lesssim10^{-3}$ (see for
example \cite{Bovy-09}), the boost factors are $\lesssim100$. 

This result suggests that additional
assumptions are needed to account for the boost needed to explain the
cosmic ray anomalies by dark matter annihilation alone. The inclusion of colder
substructures to the overall smooth component
with higher densities and lower velocity dispersions and thus, higher boost factors, could perhaps
solve the issue. However, as found in recent high-resolution N-body
simulations, the local dark matter distribution is rather smooth
\cite{Vogelsberger-09}. Typical estimates on such an additional boost factor due to 
substructure in the galactic halo (albeit without Sommerfeld enhancement)
  are of the order of 1.4, unless a subhalo happens to be very close to Earth,
in which case this boost could be larger than 10 \cite{Diemand-08}. These
estimates do not include a possible further amplification due to the scaling
of the Sommerfeld enhancement with velocity dispersion. However, the constraints found by
\cite{Slatyer-Padmanabhan-Finkbeiner-09} based on energy deposition at
recombination suggest that the enhancement must be already close to saturation
for $\sigma_v=150{\rm kms}^{-1}$, thus such a further amplification seems
implausible. Therefore, is not clear that the inclusion of colder substructures
can account for the additional boost.

Finally, we mention that we have created a web
application that solves the relevant equations described in this paper. It
allows the user to compute individual Sommerfeld boosts,
thermally averaged enhancements, cross section and $\mu$ values,
among other quantities. The interested reader can find this at
http://www.mpa-garching.mpg.de/$\sim$vogelsma/sommerfeld/ 

\section*{Acknowledgments}

JZ is supported by the Joint Postdoctoral Program in Astrophysical Cosmology of the
Max  Planck Institute for Astrophysics and the Shanghai Astronomical
Observatory. We thank the anonymous referee for helpful comments and suggestions.

\end{document}